# Same-spin correlation screens the exchange peak of the electron gas near $2k_F$

**Carl A. Kukkonen**

*33841 Mercator Isle, Dana Point, California 92629, USA*

kukkonen@cox.net

### *Abstract*

Exchange alone gives the electron gas a local field factor with a sharp peak just below $2k_F$. Quantum Monte Carlo finds none. Both spin channels are computed here from their lowest diagrams with a dynamically screened interaction and no fitting. Parallel-spin correlation is the negative image of exchange and removes about half the peak; the calculated spin-resolved factors account for the QMC data at $r_s$ = 1 and 2, and at q = 0 for the compressibility and spin susceptibility obtained from the ground-state energy. The two classes of diagrams are the two terms of the kernel of Hedin's vertex equation, of which the second is routinely dropped.

### *The puzzle*

Exchange alone makes predictions that fail. The exchange-only local field factor $G_x$ of the electron gas is known exactly [1,2]. It is universal, independent of density, and rises to 1.99 at q = 1.97 $k_F$ before falling to $\pi^2/6$ at $2k_F$, the diameter of the Fermi surface, with a singular slope. Taken alone, the exchange-only theory sharpens Friedel oscillations, deepens Kohn anomalies, and predicts spin- and charge-density-wave instabilities at $r_s \approx 5$ and 11. Screening removes these predictions, and whether any peak survives has been debated since [11,12]. Quantum Monte Carlo finds a local field factor that is smooth through $2k_F$ at every density investigated, most recently by diffusion Monte Carlo to $r_s$ = 100 [5,6,7,8,9]; Moroni, Ceperley, and Senatore looked for the predicted $2k_F$ hump in 1995 and found none [5]. The singular peak is absent and a soft crest remains. Self-consistent closures and wave-vector analyses of the response reproduce the smooth curve [3,4] and parametrizations fit it [29,30], but the cancellation itself has not been calculated.

A recent empirical decomposition by the present author identified what must be responsible [2]. Throughout, data means the variational diagrammatic Monte Carlo (VDMC) local field factors of Refs. [7,8] with their published uncertainties; diffusion Monte Carlo [5,6,9] and finite-temperature path-integral results [10] are named where used. The density and spin factors $G_+$ and $G_-$ are the sum and difference of the spin-resolved constituents, $G_{\uparrow\uparrow} = (G_+ + G_-)/2$ and $G_{\uparrow\downarrow} = (G_+ - G_-)/2$. Exchange acts only between parallel spins, so $G_{\uparrow\uparrow} = G_x + G_{\uparrow\uparrow c}$, and subtracting the exact $G_x$ removes all model error from the exchange term, leaving the parallel-spin correlation with the statistical uncertainty of the data. Correlation here means everything the Monte Carlo contains beyond first-order exchange. The result is the negative image of exchange: large, peaked at the same wavevector, opposite in sign, removing about half the peak. This Letter calculates that correlation from the Coulomb interaction and the density alone and confirms the cancellation.

### *Definitions and diagrams*

The bare interaction is $v(q) = 4\pi e^2/q^2$ and $\chi_0(q,iv)$ is the Lindhard function. The screened interaction is $W = v/(1 - v\chi_0)$, the chain of free-electron bubbles summed to all orders, with its full frequency dependence kept on the imaginary axis; no local field factor appears inside it. The local field factor is the kernel of the proper polarization, $\Pi = \chi_0/(1 + vG\chi_0)$, which is how G is extracted from the Monte Carlo response [8]. Expanding that relation, $\Pi = \chi_0 - vG\chi_0^2 + v^2G^2\chi_0^3 - \ldots$, the term linear in G is a single kernel insertion, and every term of higher order in G is two or more insertions strung together by a bare Coulomb line and is therefore reducible: cutting that line disconnects it. Each diagram of Fig. 1 is a single irreducible block, because cutting any one of its interaction lines leaves it connected. Each therefore maps onto the term linear in G alone, $-vG\chi_0^2$, where $G = -\delta\Pi/(v\chi_0^2)$ is exact rather than a linearization; the reducible chains are generated by the response function and are not extracted from the diagrams. The check is exchange: with the bare line, diagrams (a) and (b) return the exact $G_x$ of Ref. [1] as the temperature goes to zero; at the calculational temperature they agree with it to 0.2 percent at small q and fall a few percent below it near $2k_F$, a thermal deficit documented in Appendices B and D. For the spin-resolved factors the relation is applied to each element of the spin-response matrix, which the ± combinations diagonalize.

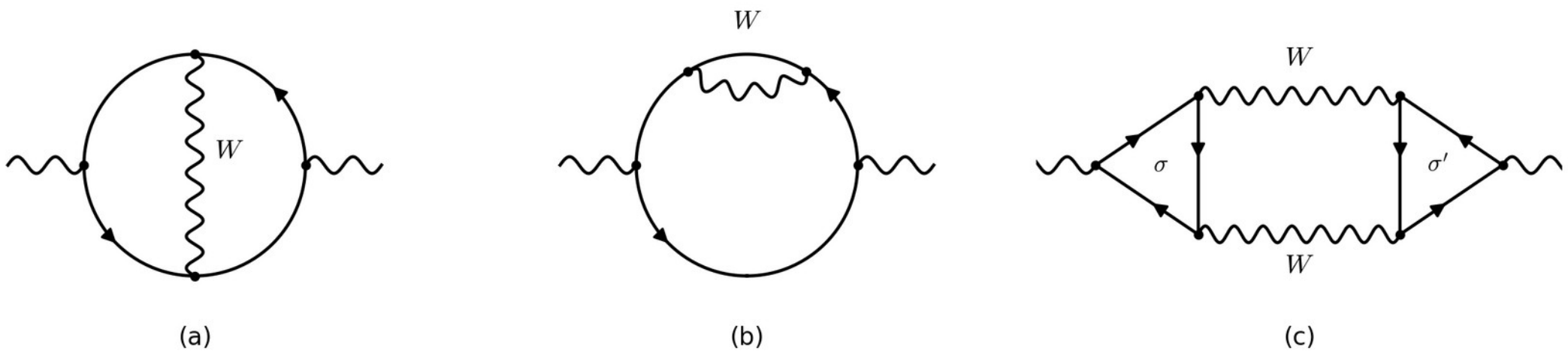

FIG. 1. The three diagrams. (a) is the first-order vertex correction and (b) the self-energy insertion that accompanies it; for any instantaneous line the two combine into a single conserving integrand [4]; with the bare line v the pair gives the exact $G_x$, with the screened line W the one-loop part of the parallel-spin channel. (c) is the lowest diagram in which opposite spins interact, two fermion triangles tied by two screened lines: one electron polarizes the medium and the other feels the induced polarization. Its partner with reversed loop orientation is included. Solid lines are bare propagators, wavy lines the interaction.

Diagrams (a) and (b) contain the interaction line exactly once and are linear in it: the diagram with W equals the diagram with v plus the diagram with W − v. The first is exact exchange, so the second is the calculated parallel-spin correlation of those diagrams. Expanding W − v in powers of v generates the infinite series of bubble-dressed exchange diagrams; at low finite order in v the $2k_F$ structure persists, and it is the summed screened line that removes it; no claim is made about every order. Diagram (c) carries two screened lines, is quadratic in W, and is a separate contribution. The set is defined by the vertex structure, the two terms of the kernel each taken once, not by order in W: the second iteration of the first term, the same-spin ladder and crossed diagrams, is omitted. Those live on a single fermion loop, so they contribute to the parallel channel only and enter $G_+$ and $G_-$ equally, leaving the cancellation of (c) from the spin channel untouched; and because they do enter the spin channel, the q = 0 spin test below bounds them at the 1 to 3 percent level. The bookkeeping is simple and is used throughout: (a) and (b) contribute to the parallel channel only, and (c) contributes the same amount to each channel, so $G_{\uparrow\uparrow}$ is (a) + (b) + (c) and $G_{\uparrow\downarrow}$ is (c) alone. Nothing is recomputed for exchange: the diagrams with the screened line give $G_{\uparrow\uparrow}$ directly, and the exact $G_x$ is subtracted only to display the correlation.

What screening does to the singular slope can be stated precisely. Engel and Vosko showed that the cubed logarithm in $G_x$ originates entirely in the small-momentum $1/k^2$ of the bare line [1]. A statically screened line has no $1/k^2$ and the singular term cancels identically. The dynamically screened line does not: at fixed frequency ν the polarization vanishes as $k^2$, so W(k,iν) retains the bare $1/k^2$ with weight $\nu^2/(\nu^2 + \omega_p^2)$, because screening fails above the plasma frequency. In the region of the integral that controls the singular behavior, where the frequency is of order the particle-hole energy, the Fermi velocity times the momentum transfer, screening is complete; the surviving weight lies above $\omega_p$ and is small. The result is a strong suppression rather than a proven exact cancellation, leaving weak nonanalytic structure at $2k_F$, which is what screened diagrammatic calculations have always shown [18] and what the data show as the soft crest. No subtraction of singular parts is performed; the diagrams are evaluated as they stand. The exact statement for the dynamic line requires the asymptotic analysis of the frequency integral and is deferred.

***The parallel-spin channel***

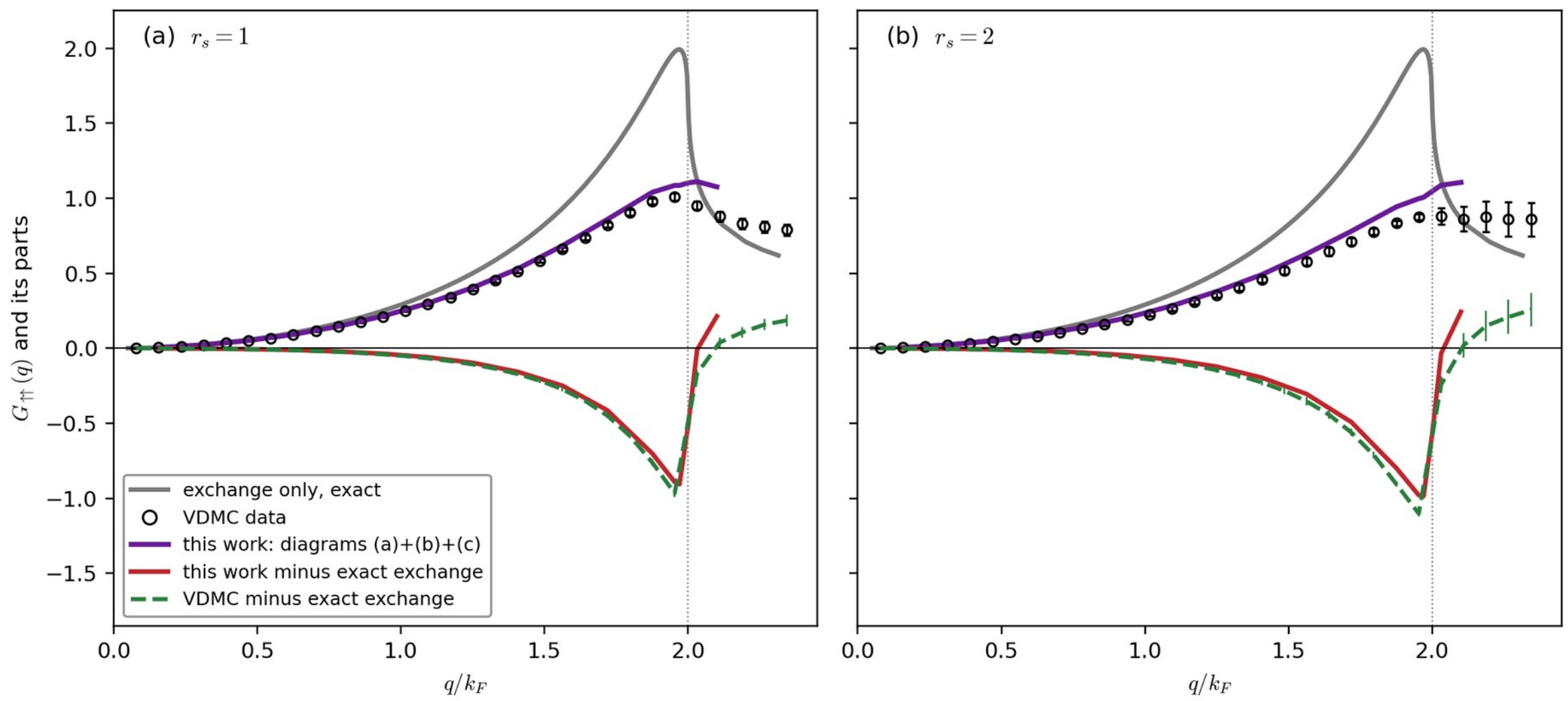

FIG. 2. The parallel-spin channel at $r_s = 1$ (a) and 2 (b), at a temperature of 0.025 of the Fermi temperature, at which the frequency integrals are done on the Matsubara axis. Gray: exact $G_x$, drawn to 2.35 $k_F$. Points: VDMC [7,8], uncertainties of $G_+$ and $G_-$ combined in quadrature assuming independence; their covariance is not published, and a positive covariance would reduce the uncertainty of $G_{\uparrow\downarrow}$ and increase that of $G_{\uparrow\uparrow}$. Purple: this work, diagrams (a), (b) and (c) with the same dynamically screened interaction line, drawn to 2.1 $k_F$, with a band of one percent for the numerical uncertainty of the calculated curves. Red: the calculated correlation of this work, the purple curve minus $G_x$. Dashed green: data minus $G_x$, the empirical correlation. The red curve closely following the dashed green curve is the primary result.

The frequency dependence of the screened interaction line is essential: over 0.16 to 1.7 $k_F$ the statically screened one-loop diagrams average 0.78 of the data $G_{\uparrow\uparrow}$ at $r_s = 1$ and the dynamic line 0.92, and the remainder is diagram (c), which contributes the same amount here as in the antiparallel channel. Diagrams (a) and (b) separately change by several times their sum when the line is made dynamic, in opposite directions, and between 86 and 93 percent of those changes cancel at $r_s = 1$ to 5 [35]; the dynamic line matters only through what survives. With it added the calculated curve lies 2.8 percent above the data at $r_s = 1$ and 7.0 percent above it at $r_s = 2$ below 1.75 $k_F$, growing toward the zone boundary. At the exchange maximum, q = 1.97 $k_F$, exchange alone reaches 1.99 while the calculated factor is 1.09 and 1.01: the correlation removes 46 and 49 percent of the peak, and the calculated curve crosses $2k_F$ with a crest of a few percent where exchange has a maximum and a cusp. Lowering the temperature fourfold raises both channels by about 3 percent below $2k_F$ and leaves them unchanged above, sharpening that crest, so the curve shown understates what survives at zero temperature rather than inventing it. Beyond 2.1 $k_F$ the diagrams are outside their domain and nothing is claimed. At q → 0 the data fall below the thermodynamic compressibility value by 3 and 7 percent, the same offsets. Diffusion Monte Carlo agrees with the VDMC density factor across this window at $r_s = 2$ [9] but has no points below 0.6 $k_F$. The origin of the offset is not established here, and until it is the data constrain omitted diagrams only at this level. Only $r_s = 1$ and 2 are shown because only there have both $G_+$ and $G_-$ been calculated by one ground-state method [8]; finite-temperature spin-resolved data exist [10]. One consequence deserves stating: both the calculation and the data keep $G_+$ within 5 percent of its long-wavelength quadratic law out to $k_F$, against nearly a factor of two for exchange alone. That is the nearly quadratic behavior the adiabatic local density approximation (ALDA) assumes, and this calculation supplies its reason. Definitions of the quoted statistics, grids, cutoffs, convergence tests, and the effect of the finite calculational temperature on the q → 0 coefficients, which is below 0.2 percent, are in Appendices C to F.

### *The physical picture*

The Pauli principle digs a like-spin hole around each electron. Because the Fermi surface is sharp, the hole has long-range Friedel oscillations at wavevector $2k_F$. First-order exchange is the electron interacting with this hole through the bare Coulomb force, which is also long-range and reaches every oscillation, and the contributions add coherently over many Fermi wavelengths. That coherent sum is the $2k_F$ peak. Now turn on the electron-electron interaction. Parallel spins repel a little more, but Pauli has already emptied the neighborhood, and that effect is short range and modest. The decisive actor is the other spin population. The parallel-spin correlation being calculated is the change in the same-spin response; the agent that produces it is the screening cloud, which both spin populations form. The antiparallel electrons are the new element: Pauli never touched them, and the Coulomb interaction displaces them for the first time. Every electron gathers a screening cloud that neutralizes its field beyond a screening length, less than half an oscillation period at $r_s = 1$. The distant parallel-spin partner no longer feels a bare charge. The oscillations remain, but the force no longer reaches them, and the peak is cut down. This is consistent with diagrams (a) and (b) being same-spin physics: their external vertices carry one spin, and the other spin enters inside the screened line, because the screening cloud is built from both populations.

***The antiparallel-spin channel and the test at q = 0***

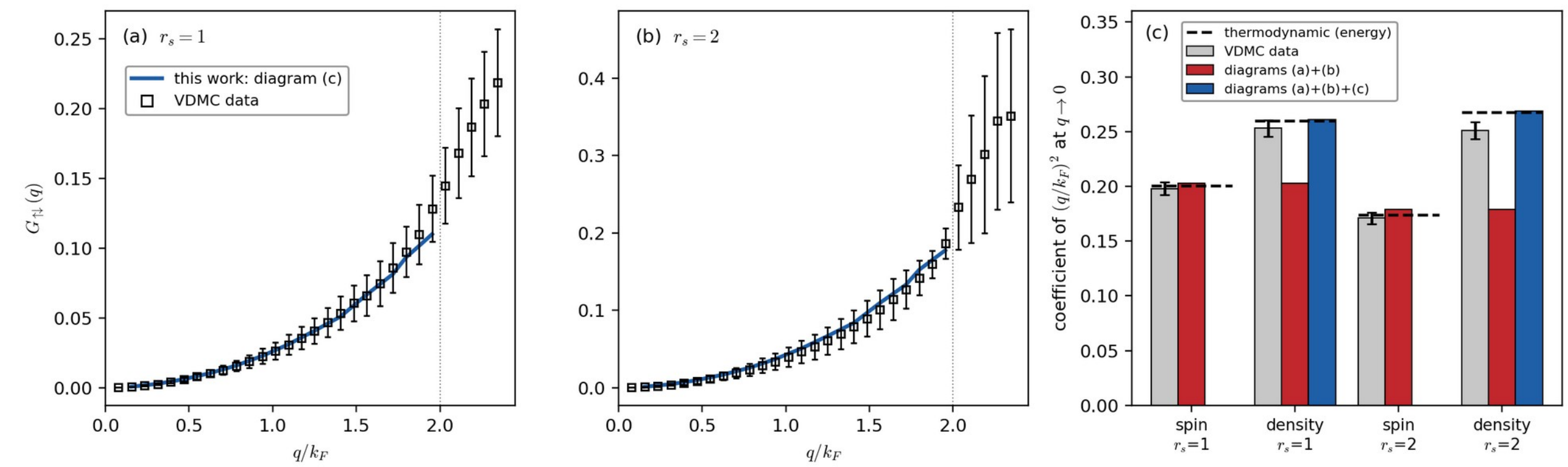

FIG. 3. (a), (b): the antiparallel-spin channel $G_{\uparrow\downarrow}$ at $r_s = 1$ and 2. Points: VDMC, uncertainties as in Fig. 2; because $G_{\uparrow\downarrow}$ is half the difference of two nearly equal quantities its relative uncertainty is 20 to 30 percent. Blue: diagram (c) with the dynamically screened line. (c): the coefficient of $(q/k_F)^2$ at $q \to 0$ from the data, from diagrams (a) and (b) alone, and, in the density channel, with (c) added. Dashed: the thermodynamic values from the ground-state energy. Values in the text.

In the irreducible kernel, opposite spins first interact at second order in W: spin is conserved along a fermion loop, and a single interaction line cannot put the two external vertices on separate loops without making the diagram reducible. In diagram (c) the external vertices sit on different loops, and with a spin-blind interaction the amplitude is the same whether the two loop spins are parallel or antiparallel; the diagram therefore contributes equally to $G_{\uparrow\downarrow}$ and $G_{\uparrow\uparrow}$ and cancels from the spin channel exactly. Within the kernel considered here it is the only place opposite spins enter; no claim is made about the exact vertex. With the dynamic line it reproduces $G_{\uparrow\downarrow}$ with mean ratios 1.00 and 1.09, inside the data's 20 to 30 percent uncertainty, a consistency test rather than a sharp one; with a statically screened line the ratios are 0.82 and 0.84. The diagram has been evaluated before in the effective-interaction literature [14,15,18]. In three dimensions $G_-$ has been supplied by model interpolations for two decades; Simion and Giuliani identified its unknown behavior at intermediate wavevector as the factor limiting the accuracy of the effective mass and spin susceptibility, and asked for Monte Carlo to settle it [28]. That is the object computed here.

At q = 0 the answers are known independently, channel by channel, from thermodynamics: the compressibility fixes the leading coefficient of $G_+$ and the spin susceptibility that of $G_-$, both from the Perdew-Wang correlation energy and its spin stiffness [20,21], with no local field factor used. The long-wavelength limits of the calculated curves are computed, not imposed, so the comparison is a test rather than a check of self-consistency. Within the RPA screening used here, a spin field does not move the screening, because W depends on the total density only, so the spin channel tests diagrams (a) and (b) alone and diagram (c) is absent from it exactly; a density field does move the screening, so the compressibility tests all three. That these three diagrams together are the second density derivative of the exchange plus ring-diagram energy at $q \to 0$ was shown by Geldart and Vosko [34]; the test here is its spin-resolved form against the full correlation energy. The numbers, for $r_s = 1$ and 2 in that order, all taken from the Perdew-Wang parametrization [20,21] by the procedure given in Appendix E: in the spin channel the first-order coefficient is 0.2030 and 0.1790 against thermodynamic 0.2005 and 0.1737, high by 1.2 and 3.1 percent; diagram (c) alone gives 0.0290 and 0.0450 against the thermodynamic antiparallel coefficient 0.0297 and 0.0470, low by 2.5 and 4.3 percent; in the density channel, where (c) enters twice, the sum is 0.2610 and 0.2690 against 0.2600 and 0.2677, high by 0.4 and 0.5 percent. Each class of diagram accounts for its own sum rule to a few percent; the density channel agrees to half a percent, better than either piece alone, because the first-order coefficient is slightly high and the two-triangle coefficient slightly low and the two deviations nearly cancel in the sum. Measured instead against the exchange plus RPA correlation energy, of which these diagrams are the exact second derivative, the agreement is 0.2 percent in the spin channel and better than 0.1 percent in the density channel at $r_s$ = 1, 2 and 5 [35]. The one-to-three percent offsets against the full correlation energy are therefore the correlation the ring diagrams omit, not numerical error in the diagrams. The variation of the screening, diagram (c), is the single mechanism that supplies the antiparallel channel and the long-wavelength shortfall of the parallel one. The quasiparticle weight cannot supply the missing piece: in the Fermi-liquid expression for the compressibility it enters through two limits of the vertex function and cancels between them [23].

***The two classes of diagrams and Hedin's equations***

The vertex solves an integral equation whose kernel is the derivative of the self-energy with respect to the propagator [22]. With the self-energy in the GW form, the product of propagator and screened interaction, that derivative has two contributions, one from the explicit propagator and one from the propagator implicit in the screening: the screened interaction itself, and the variation of the screening, which at this order is a pair of diagrams. The first, iterated once, gives Fig. 1(a) with its self-energy partner (b) and carries same-spin physics only, because the electron and hole of one bubble share a fermion loop. The second gives Fig. 1(c), whose second loop carries either spin. The second contribution is dropped in the usual Bethe-Salpeter kernel [33,24] and in the first-order vertex, retained in some self-consistent schemes, or replaced by a local kernel [23,25,26]. Where it is retained, its diagrams are known to permit spin flips and to improve the calculated dielectric function of the electron gas, and among the schemes studied there, only vertices containing it reproduce the pole in the compressibility near $r_s$ = 5.25 [23]. The two-channel effective interaction of 1979 contained both contributions by construction [16,17], and the same conclusion was drawn in local field language twenty years ago: a vertex carrying only charge fluctuations has no antiparallel-spin content, which resides in $G_-$ [27,28]. What has not been done before is to resolve this contribution into its two spin channels, evaluate each, and test them against spin-resolved data. Together the two classes of diagrams give a spin-resolved exchange-correlation kernel, $f_{xc}(q) = -4\pi e^2 G_+/q^2$, with no adjustable parameter, tested here at two metallic densities below $2k_F$; the same diagrams give its frequency dependence [31,32].

***What is new***

The diagrams are old. Geldart and Vosko evaluated all three with the dynamically screened line in 1966, showed that the two-triangle diagram is as large as the other two at metallic density, and showed that together they give the compressibility of the ring-diagram energy at $q \to 0$ [34]. First-order dynamic exchange was studied in 1976 and 1979 [12,13], the two-triangle diagram reappears in the effective-interaction literature of 1986 [14], and Richardson and Ashcroft evaluated the screened versions of all three in 1994 [18], as did the conserving treatments of the period [19]. None was compared with spin-resolved data, because none existed. The two first-order diagrams have since been evaluated in two dimensions with a screened interaction [36]. Their combination into a single conserving integrand, and how much of it survives a retarded interaction, is the subject of Ref. [35]. What is new is the spin resolution and the test: exchange is subtracted from spin-resolved data with no model error, the two classes of diagrams are compared with it channel by channel, and each accounts for its own long-wavelength sum rule. The electron gas shows no exchange peak because parallel-spin correlation is exchange's own screening: the negative image of exchange, of which only the sum had been seen.

***Appendices***

Appendices A to H give the derivation of the local field factor from a diagram (A), the recovery of exact exchange (B), the numerical method and convergence (C), the effect of temperature (D), the long-wavelength coefficients (E), the definitions of the quoted statistics (F), the tables of the calculated curves (G), and the verification record (H). Numbers are in atomic units with momenta in units of kF unless stated. The Monte Carlo data are the variational diagrammatic Monte Carlo local field factors of Refs. [7,8].

***Appendix A. The local field factor from a diagram***

The local field factor is defined as the kernel of the proper polarization, $\Pi = \chi_0/(1 + vG\chi_0)$, where $\chi_0$ is the Lindhard function and $v = 4\pi e^2/q^2$. This is also the relation by which G is extracted from the Monte Carlo response [8]. Expanding it in powers of the kernel gives

$$\Pi = \chi_0 - vG\chi_0^2 + v^2G^2\chi_0^3 - \dots .$$

The first term, $\chi_0$, contains no kernel. The term of order n in G is n kernel insertions joined to one another by n − 1 bare Coulomb lines. The term linear in G, $-vG\chi_0^2$, is a single insertion; every term of higher order in G is reducible, because cutting one of its connecting Coulomb lines disconnects it. Each diagram of Fig. 1 is a single block that remains connected when any one of its interaction lines is cut. Diagrams (a) and (b) have one screened line and do not fall apart when it is cut, because the fermion loop still joins the two external vertices; diagram (c) has two screened lines and remains connected when either is cut. Each diagram is therefore a contribution δΠ to the term linear in G alone, and its local field factor follows from that term:

$G = -\delta\Pi/(v\chi_0^2)$.

This is exact for an irreducible insertion and is not a linearization. The reducible chains of the higher terms are generated when the kernel is inserted into the response function; they are properties of the response, not of the kernel. Extracting a first-order kernel through the full nonlinear inversion of Π would count those chains once as absent and once as present. The check of the normalization is Appendix B. The check of the irreducibility argument itself cannot be made at first order in v, where the linear and nonlinear relations coincide, and none is claimed.

Spin resolution. With a spin-independent Coulomb interaction and the local field kernel written as the spin matrix $G_{\sigma\sigma'}$, the proper spin-response matrix obeys $\Pi^{-1} = (2/\chi_0)I + 2vG$, where I is the unit matrix, $\chi_0$ is the total Lindhard function and $\chi_0/2$ the per-spin one. In the unpolarized gas $G_{\uparrow\uparrow} = G_{\downarrow\downarrow}$ and $G_{\uparrow\downarrow} = G_{\downarrow\uparrow}$, and the matrix has eigenvectors (1, 1) and (1, −1). Along them

$\Pi_s^{-1} = 1/\chi_0 + v(G_{\uparrow\uparrow} + G_{\uparrow\downarrow}) = 1/\chi_0 + vG_+$,   $\Pi_a^{-1} = 1/\chi_0 + v(G_{\uparrow\uparrow} - G_{\uparrow\downarrow}) = 1/\chi_0 + vG_-$,

so $G_+ = G_{\uparrow\uparrow} + G_{\uparrow\downarrow}$ and $G_- = G_{\uparrow\uparrow} - G_{\uparrow\downarrow}$ hold exactly, which is the normalization of Ref. [8]. The corresponding linear extraction for a single spin element is $G_{\sigma\sigma'} = -2\delta\Pi_{\sigma\sigma'}/(v\chi_0^2)$. A diagram on one fermion loop, such as (a) or (b), has $\delta\Pi_{\sigma\sigma'} = \delta_{\sigma\sigma'}P$ and contributes to $G_{\uparrow\uparrow}$ only, hence equally to $G_+$ and $G_-$. Diagram (c) has its external vertices on two different loops joined by spin-blind lines; fixing the two external spins fixes the two loop spins, and nothing in the integrand depends on whether they are equal, so $\delta\Pi_{\sigma\sigma'} = A$ for all four spin pairs. It therefore contributes A to both $G_{\uparrow\uparrow}$ and $G_{\uparrow\downarrow}$, 2A to $G_+$, and nothing to $G_-$.

***Appendix B. Recovery of exact exchange***

With the bare line v in place of W, diagrams (a) and (b) are first-order exchange and should return the exact $G_x$ of Engel and Vosko [1]. They were evaluated with the production code at $r_s = 1$ on the q grid of the Monte Carlo data and compared with the closed form. The table gives the result at the production temperature $T = 0.025\ T_F$.

| **$q/k_F$** | **diagrams (a)+(b), bare line** | **exact $G_x$** | **deviation (%)** |
|---|---|---|---|
| 0.1562 | 0.00616 | 0.00621 | -0.73 |
| 0.3125 | 0.02488 | 0.02485 | +0.12 |
| 0.4688 | 0.05685 | 0.05679 | +0.09 |
| 0.6250 | 0.10341 | 0.10341 | +0.00 |
| 0.7812 | 0.16656 | 0.16692 | -0.21 |
| 0.9375 | 0.24975 | 0.25069 | -0.37 |
| 1.0938 | 0.35758 | 0.35957 | -0.55 |
| 1.2500 | 0.49559 | 0.50083 | -1.05 |
| 1.4063 | 0.67577 | 0.68604 | -1.50 |
| 1.5625 | 0.91572 | 0.93317 | -1.87 |
| 1.7187 | 1.23724 | 1.27240 | -2.76 |
| 1.8750 | 1.65712 | 1.74055 | -4.79 |

At small q the agreement is within 0.2 percent after the first point. Toward $2k_F$ a deficit grows, reaching 4.8 percent at 1.875 $k_F$. It is thermal. The same diagrams were recomputed at lower temperature at two wavevectors:

| **$q/k_F$** | **$T = 0.025\ T_F$** | **$T = 0.0125\ T_F$** | **$T = 0.00625\ T_F$** | **extrapolated to T = 0** |
|---|---|---|---|---|
| 1.4063 | -1.50 % | -0.90 % | -0.33 % | 0.25 % |
| 1.8750 | -4.79 % | -2.74 % | -1.29 % | 0.15 % |

The deficit roughly halves each time the temperature halves and extrapolates linearly to zero within a quarter of a percent. The diagrams therefore recover exact exchange as T → 0, and at the production temperature they carry a thermal deficit near $2k_F$: about 1.5 percent at 1.4 $k_F$ and 5 percent at 1.9 $k_F$. The calculated curves share this deficit, which is the quantitative basis for the statement that the finite-temperature curve understates the structure near $2k_F$. The deficit is common to the bare and screened evaluations and largely cancels in their difference, the calculated correlation of the linearity identity.

***Appendix C. Numerical method and convergence***

Parallel channel, diagrams (a) and (b). Bare propagators; dynamically screened RPA line $W(p, i\nu) = v/(1 - v\chi_0)$ on the imaginary axis; Matsubara sum over bosonic frequencies $\nu_l = 2\pi l/\beta$ to $l = L_{max} = 1600$ at $T = 0.025\ T_F$, with $L_{max}$ scaled as 1/T at lower temperature so that the frequency cutoff is fixed; the chemical-potential counterterm holds the density fixed. Momentum grids: outer Nk = 12, Nμ = 12, inner $Nk_2 = 16$, $N\mu_2 = 20$, Nφ = 12, nτ = 8.

Antiparallel channel, diagram (c). Two fermion triangles, each evaluated by partial fractions over the internal fermion frequency, joined by two dynamically screened lines; the bosonic frequency is integrated continuously by Gauss-Legendre quadrature on $\nu = t/(1 - t)$ with 14 points. The triangle grid is spherical about the origin with radial points clustered at |k| = 1. Production and extended grids: triangle 40×32×20 with kmax 2.6 and p-grid 16×10 with pmax 4; extended 48×32×20, kmax 6, p-grid 32×10, pmax 8. All values in the Letter use the extended grid.

Reproduction. Each code, restructured for the temperature and window studies, reproduces its stored production values to better than 0.0003 percent at every point checked.

Antiparallel channel below the window. Joint refinement of every triangle and p parameter, four levels, at q = 0.156, 0.234 and 0.313; largest step in the integral across the three wavevectors:

| **step** | **static $r_s = 1$** | **static $r_s = 2$** | **dynamic $r_s = 1$** | **dynamic $r_s = 2$** |
|---|---|---|---|---|
| L0 → L1 | 1.02 % | 1.08 % | 1.07 % | 1.10 % |
| L1 → L2 | 0.43 % | 0.34 % | 0.52 % | 0.45 % |
| L2 → L3 | 0.15 % | 0.15 % | 0.18 % | 0.18 % |

At finite q, doubling the momentum extents at matched resolution moves the value at q = 1.09 $k_F$ by 1.0 percent; doubling the frequency points moves it by 0.001 percent. The tabulated curves carry a numerical uncertainty of 1 percent below 1.9 $k_F$ on this basis.

Antiparallel channel in the window, q = 1.953, 1.970 and 2.000 $k_F$, the same four levels, dynamic line; largest step across the three wavevectors and both densities:

| **T / $T_F$** | **L0 → L1** | **L1 → L2** | **L2 → L3** |
|---|---|---|---|
| 0.025 | 1.65 % | 1.10 % | 0.70 % |
| 0.0125 | 2.69 % | 1.59 % | 1.22 % |
| 0.00625 | 4.44 % | 2.20 % | 1.66 % |

Limitation, stated without qualification. In the window the triangle integral does not converge to half a percent on this family of grids at any temperature. The step shrinks by a factor of about 0.75 per level while each level costs four times the last, so further uniform refinement is not a route to convergence. The values oscillate between levels, and the sign is systematic in q: 1.953 and 1.970 move together while 2.000 moves opposite, at every level and temperature, identically for the static and dynamic lines and for both densities. The residual is therefore a property of the triangle integral and not of the line. The Letter uses the midrange of the top three levels at the production temperature, for which half the spread is 0.34 percent at 1.953 and 1.970 $k_F$ and 0.53 percent at 2.000 $k_F$; this is a spread, not an error bar. The antiparallel diagram contributes about one sixth of the parallel curve in the window, so its contribution to the one percent band of Fig. 2 is below 0.2 percent. The grid is not adapted to the Fermi surfaces of the second and third propagators, which are displaced from the origin, and a boundary-adapted grid is the expected remedy; it has not been tested.

Parallel channel, frequency cutoff and time grid: doubling $L_{max}$ changes the value at 1.97 $k_F$ by $2\times10^{-7}$ percent and raising the τ-panel points from 8 to 12 changes it by 0.02 percent. Limitation: the momentum cutoffs of this channel were audited at T = 0.025 $T_F$ only.

***Appendix D. Temperature***

The calculation runs at T = 0.025 $T_F$, at which the frequency integrals are done on the Matsubara axis; the thermodynamic references of Appendix E are ground-state values. Three effects are separated.

At $q \to 0$, rebuilding the independent route-2 integral of Appendix E at T = 0.025 $T_F$ shifts the two-triangle coefficient by −0.1 percent relative to T = 0, for both the static and dynamic lines. The effect of the finite temperature on the long-wavelength coefficients is therefore below 0.2 percent.

At an interior wavevector, q = 1.406 $k_F$, lowering the temperature from 0.025 to 0.00625 $T_F$ changes the parallel channel by +0.21 and +0.22 percent at $r_s$ = 1 and 2, and the antiparallel channel by −1.3 percent. That drift is entirely in the triangles' Fermi factors; the dynamic line is flat in temperature to better than 0.01 percent.

In the window the effect is larger and has a definite shape. At fixed grid level, lowering the temperature from 0.025 to 0.00625 $T_F$ raises both channels by about 3 percent at 1.95 and 1.97 $k_F$ and leaves them unchanged at 2.00 $k_F$. The parallel channel was measured independently of the antiparallel one and shows the same pattern. Lowering the temperature therefore sharpens the crest at $2k_F$ rather than removing it. Together with the exchange deficit of Appendix B, which has the same sign and a comparable magnitude, this means the curves shown understate the structure that survives at zero temperature. No extrapolation to zero temperature is used in the Letter, because in the window the antiparallel grid spread at the lower temperatures is about a third of the temperature effect.

***Appendix E. The long-wavelength coefficients***

Thermodynamic values. The leading coefficient of $G_+$ in $(q/k_F)^2$ is fixed by the compressibility and that of $G_-$ by the spin susceptibility. Writing the sum rules as $G_+ \to A_p(q/k_F)^2$ and $G_- \to A_m(q/k_F)^2$, one has $A_p = -(k_F^2/4\pi)\partial^2(n\varepsilon_{xc})/\partial n^2$, the long-wavelength exchange-correlation kernel in units of the Coulomb interaction, and $A_m = -(3\pi/4k_F)\partial^2\varepsilon_{xc}/\partial\zeta^2$ at $\zeta = 0$, both evaluated from the Perdew-Wang parametrization of the correlation energy and its spin stiffness [20,21]; the exchange parts are exact. The antiparallel coefficient is $(A_p - A_m)/2$. The values were evaluated with the published implementation accompanying Ref. [30].

Calculated values. The first-order coefficient $c_1$ and the two-triangle coefficient a are obtained by three-point extrapolation in $q^2$ through q = 0.156, 0.234 and 0.313 $k_F$, the convention also used for the data. The two-triangle coefficient was also computed independently from its $q \to 0$ limit, $A(0) = -½ \int d\nu/2\pi \int d^3p/(2\pi)^3\, W(p, i\nu)^2\, [\partial\chi_{0\sigma}(p, i\nu)/\partial\mu]^2$, with the Fermi-surface derivative in closed form and no code shared with the production path. The extrapolated values are 0.0289 and 0.0452 at $r_s$ = 1 and 2 with the momentum cutoff of the production code; adding the tail beyond that cutoff, known from the closed form, gives 0.0290 and 0.0454, against 0.0292 and 0.0456 from the independent integral, agreement within 0.4 percent.

| channel | $r_s$ | calculated | thermodynamic | deviation |
|---|---|---|---|---|
| spin, $G_-$ | 1 | $c_1$ = 0.2030 | 0.2005 | +1.2 % |
| spin, $G_-$ | 2 | $c_1$ = 0.1790 | 0.1737 | +3.1 % |
| antiparallel | 1 | a = 0.0290 | 0.0297 | −2.5 % |
| antiparallel | 2 | a = 0.0450 | 0.0470 | −4.3 % |
| density, $G_+$ | 1 | $c_1$ + 2a = 0.2610 | 0.2600 | +0.4 % |
| density, $G_+$ | 2 | $c_1$ + 2a = 0.2690 | 0.2677 | +0.5 % |

The first-order coefficient is the same in both channels because diagrams (a) and (b) contribute to $G_{\uparrow\uparrow}$ only. Diagram (c) is absent from the spin channel exactly, within the RPA screening used, because a spin field does not change the total-density screening. The density-channel agreement is closer than either component because $c_1$ is high and a is low. The Monte Carlo coefficients, by the same extrapolation, are 0.253 and 0.251 in the density channel and 0.198 and 0.171 in the spin channel, with an extrapolation uncertainty of about 3 percent. The spin stiffness is the less certain of the two thermodynamic references, at a few percent.

Comparison with the RPA energy. The three diagrams with the screened line are the exact second density derivative of the exchange plus RPA (ring-diagram) correlation energy at $q \to 0$, an identification due to Geldart and Vosko (Ref. [34]). Measured against that energy instead of the full Perdew-Wang correlation energy, the comparison closes. With the same code, and coefficients read by successive-point solves at $q = 0.1$, 0.2 and 0.4 $k_F$ with no fit, the first-order coefficient is 0.2032, 0.1793 and 0.1394 at $r_s = 1$, 2 and 5, against 0.2027, 0.1790 and 0.1392 from the RPA spin stiffness, high by 0.2 percent; and the density-channel sum, with the two-triangle coefficient from the computed graphs, is 0.2612, 0.2701 and 0.2919, against 0.2609, 0.2700 and 0.2921 from the RPA correlation energy, agreement better than 0.1 percent (Ref. [35]). These first-order coefficients differ from those of the table above in the fourth digit only, because the extrapolation points differ. The one-to-three percent offsets in the table above are therefore the correlation that the ring diagrams omit, not numerical error in the diagrams.

Dynamic corrections diagram by diagram. When the statically screened line is replaced by the dynamic one, diagrams (a) and (b) each change by seven to ten times the change in their sum, in opposite directions; the fraction of the vertex diagram's change that the self-energy diagram cancels is 0.86 to 0.93 over 0.1 to 1.9 $k_F$ at $r_s = 1$, 2, 3 and 5, and the net increase of the one-loop sum is 14 to 62 percent, rising monotonically with $q$ and $r_s$ (Ref. [35]). The split between (a) and (b) depends on the chemical-potential counterterm convention, here fixed density; the sum does not.

### *Appendix F. Definitions of the quoted statistics*

A quoted ratio is the unweighted arithmetic mean of the pointwise ratio calculation/data over the stated interval, taken at the q points of the Monte Carlo grid of Ref. [8]. A quoted offset is that mean minus one. No weighting by uncertainty is applied, because the Monte Carlo uncertainties of $G_+$ and $G_-$ are published without their covariance, which the weights would require. Uncertainties of $G_{\uparrow\uparrow}$ and $G_{\uparrow\downarrow}$ are formed as half the quadrature sum of those of $G_+$ and $G_-$, assuming independence; a positive covariance would reduce the uncertainty of $G_{\uparrow\downarrow}$ and increase that of $G_{\uparrow\uparrow}$. The values recomputed from the stored tables are:

| statistic | interval ($k_F$) | points | $r_s$ = 1 | $r_s$ = 2 |
|---|---|---|---|---|
| one-loop, static line / data | 0.16–1.72 | 11 | 0.780 | 0.692 |
| one-loop, dynamic line / data | 0.16–1.72 | 11 | 0.917 | 0.874 |
| diagrams (a)+(b)+(c) / data | 0.16–1.72 | 11 | 1.028 | 1.070 |
| diagram (c) / data, antiparallel | 0.16–1.88 | 13 | 0.995 | 1.091 |
| diagrams (a)+(b)+(c) / data, at 1.97 | single point | 1 | 1.089 | 1.149 |

The correlation removed at the exchange maximum, q = 1.97 $k_F$, is $1 - G_{\uparrow\uparrow}/G_x$ = 46 and 49 percent at $r_s = 1$ and 2.

### *Appendix G. Tables of the calculated curves*

Dynamically screened line, T = 0.025 $T_F$, extended grid. Columns: diagrams (a)+(b); diagram (c), which is $G_{\uparrow\downarrow}$ and also its equal contribution to $G_{\uparrow\uparrow}$; their sum $G_{\uparrow\uparrow}$; exact $G_x$; and the calculated correlation $G_{\uparrow\uparrow} - G_x$. Numerical uncertainty 1 percent below 1.9 $k_F$, larger in the window as described in Appendix C; the thermal deficit of Appendix B and Appendix D applies near $2k_F$. Values above 2.1 $k_F$ are outside the domain of the diagrams and are given for completeness only.

**$r_s$ = 1:**

| $q/k_F$ | (a)+(b) | (c) = $G_{\uparrow\downarrow}$ | $G_{\uparrow\uparrow}$ | $G_x$ | $G_{\uparrow\uparrow} - G_x$ |
|---|---|---|---|---|---|
| 0.1562 | 0.00496 | 0.00070 | 0.00566 | 0.00621 | -0.00055 |
| 0.3125 | 0.01997 | 0.00275 | 0.02272 | 0.02485 | -0.00213 |
| 0.4688 | 0.04539 | 0.00602 | 0.05141 | 0.05679 | -0.00539 |
| 0.6250 | 0.08182 | 0.01038 | 0.09220 | 0.10341 | -0.01121 |
| 0.7812 | 0.13026 | 0.01579 | 0.14605 | 0.16692 | -0.02087 |
| 0.9375 | 0.19211 | 0.02271 | 0.21482 | 0.25069 | -0.03587 |
| 1.0938 | 0.26926 | 0.03100 | 0.30027 | 0.35957 | -0.05930 |
| 1.2500 | 0.36371 | 0.04076 | 0.40447 | 0.50083 | -0.09636 |
| 1.4063 | 0.47879 | 0.05098 | 0.52976 | 0.68604 | -0.15628 |
| 1.5625 | 0.61723 | 0.06671 | 0.68393 | 0.93317 | -0.24924 |
| 1.7187 | 0.77773 | 0.08137 | 0.85910 | 1.27240 | -0.41330 |
| 1.8750 | 0.93844 | 0.10152 | 1.03996 | 1.74055 | -0.70060 |
| 1.9531 | 0.97564 | 0.10995 | 1.08560 | 1.97635 | -0.89076 |
| 1.9700 | 0.97283 | 0.11289 | 1.08573 | 1.99172 | -0.90599 |
| 2.0000 | 0.98026 | 0.12244 | 1.10270 | 1.61821 | -0.51551 |
| 2.0312 | 0.97702 | 0.13418 | 1.11120 | 1.12083 | -0.00964 |
| 2.1000 | 0.92733 | 0.14873 | 1.07606 | 0.86259 | 0.21348 |
| 2.1875 | 0.90139 | 0.16361 | 1.06500 | 0.72705 | 0.33794 |
| 2.2656 | 0.89096 | 0.17397 | 1.06493 | 0.65222 | 0.41271 |

$r_s$ **= 2:**

| **q**/$k_F$ | **(a)+(b)** | **(c) =** $G_{\uparrow\downarrow}$ | $G_{\uparrow\uparrow}$ | $G_x$ | $G_{\uparrow\uparrow} - G_x$ |
|---|---|---|---|---|---|
| 0.1562 | 0.00438 | 0.00110 | 0.00547 | 0.00621 | -0.00074 |
| 0.3125 | 0.01760 | 0.00435 | 0.02195 | 0.02485 | -0.00291 |
| 0.4688 | 0.03992 | 0.00960 | 0.04953 | 0.05679 | -0.00727 |
| 0.6250 | 0.07178 | 0.01669 | 0.08847 | 0.10341 | -0.01494 |
| 0.7812 | 0.11386 | 0.02552 | 0.13938 | 0.16692 | -0.02754 |
| 0.9375 | 0.16710 | 0.03689 | 0.20399 | 0.25069 | -0.04669 |
| 1.0938 | 0.23277 | 0.05056 | 0.28333 | 0.35957 | -0.07624 |
| 1.2500 | 0.31213 | 0.06666 | 0.37879 | 0.50083 | -0.12204 |
| 1.4063 | 0.40722 | 0.08350 | 0.49072 | 0.68604 | -0.19532 |
| 1.5625 | 0.51934 | 0.10937 | 0.62871 | 0.93317 | -0.30446 |
| 1.7187 | 0.64739 | 0.13303 | 0.78042 | 1.27240 | -0.49197 |
| 1.8750 | 0.77785 | 0.16478 | 0.94264 | 1.74055 | -0.79792 |
| 1.9531 | 0.82009 | 0.17734 | 0.99743 | 1.97635 | -0.97892 |
| 1.9700 | 0.82464 | 0.18175 | 1.00639 | 1.99172 | -0.98533 |
| 2.0000 | 0.85014 | 0.19652 | 1.04667 | 1.61821 | -0.57154 |
| 2.0312 | 0.87271 | 0.21443 | 1.08715 | 1.12083 | -0.03369 |
| 2.1000 | 0.87012 | 0.23559 | 1.10572 | 0.86259 | 0.24313 |
| 2.1875 | 0.88255 | 0.25793 | 1.14048 | 0.72705 | 0.41343 |
| 2.2656 | 0.89977 | 0.27446 | 1.17423 | 0.65222 | 0.52201 |

***Appendix H. Verification record***

This record states the tool, its role, the direction given, and the means of verification. The work used Claude (Anthropic; Claude Fable 5 and Opus models, through claude.ai, 2026) for literature search and verification, derivations, numerical code, figures, and drafting. The author directed every step through written task briefs, each specifying the calculation, the checks it had to pass before any number could be reported, and the instruction to stop and report rather than adjust when a check failed. Derivations were carried out one step at a time, with the author accepting or challenging each step before the next.

Verification by independent routes. Each principal number was checked by a route sharing no code with the production calculation: the two-triangle long-wavelength coefficient by the closed-form integral of Appendix E; the normalization of diagrams (a) and (b) by the recovery of exact exchange in Appendix B; the thermodynamic references by an independent published implementation. Every restructured code was required to reproduce its stored values before new values were accepted.

Gated calculations, including those that failed. The dynamic-line two-triangle calculation first failed its small-q gate, was stopped, and was completed only after a joint grid refinement and a change to the extrapolation convention agreed in advance. The temperature study through the window stopped on two failed gates and was reported as ungated. The grid ladder in the window did not meet its convergence criterion and is reported as a spread (Appendix C). In each case the task reported the failure rather than adjusting settings to pass.

Corrections made before submission. The process caught errors of substance, which are recorded here because they bear on how much the result should be trusted. The two-triangle diagram was at first assigned to the antiparallel channel alone; a spin-algebra derivation showed that it contributes equally to both channels, which changed the parallel-channel curve and the long-wavelength analysis. A second-order calculation that had appeared to account for the long-wavelength deficit was shown, by testing the two channels separately, to have the wrong channel structure, and was withdrawn. An early estimate of the sign of the occupancy contribution above $2k_F$ was wrong and was corrected against the bookkeeping of Richardson and Ashcroft. The long-wavelength percentages were at one stage computed against inconsistent reference values and were recomputed from a single source (Appendix E). Two quoted statistics were corrected when recomputed for Appendix F.

Literature. Roughly two hundred specific statements about prior work were checked against the original papers, including those supplied by three independent searches with other artificial intelligence systems; citations were verified individually, and several attributions found in secondary sources were corrected. Three independent referee-style reads of the manuscript by artificial intelligence systems were obtained before submission and all points raised were addressed.

The author takes full responsibility for the content.

### *Data availability*

The calculated curves are tabulated in Appendix G. The Monte Carlo data are those of Refs. [7,8].

### *Acknowledgments*

The author conceived the subtraction of exact exchange from the spin-resolved Monte Carlo data, carried out that decomposition in Ref. [2], chose the diagrams and the checks each result had to pass, and developed the physical picture of correlation as the screening of exchange. AI tools (Anthropic Claude, 2026) assisted, under the author's written task briefs with the checks set in advance, with literature search and verification, derivations, numerical code, figures, and drafting. Every citation was verified against the original paper and every number against exact limits and an independent code path, every derivation step was reviewed by the author, and the author is solely responsible for the paper. The verification record is Appendix H.